\documentstyle[aps]{revtex}
\begin{document}

\title{QUATERNION ANALYSIS}
\author{
Khaled Abdel-Khalek\footnote{Work supported by an
ICSC--World Laboratory scholarship.\\
e-mail : khaled@le.infn.it}
}
\address{ 
Dipartimento di Fisica - Universit\`a di Lecce\\
- Lecce, 73100, Italy -}

\date{ 1996}
\maketitle
\abstract{Quaternion analysis is considered in full details 
where a new analyticity condition in complete analogy to
complex analysis is found. 
The extension to octonions
is also worked out.}

\widetext
\section{Introduction}
The ultimate goal of theoretical physics is to understand 
nature at any scale. Whatever this task seems too ambitious, 
we have already made some success. Guided by the gauge 
symmetry principle, we reached the standard model. Well 
tested but not well understood. It contains many puzzles 
\cite{zee}. 
We can go beyond the standard model by enlarging the gauge 
symmetry \cite{zee}.  Some of the standard model 
problems can be 
solved but many remain. Unifying gravity with the other 
forces of nature is still a challenge. Having a large 
symmetry in our model is a good guide. Supersymmetry and 
supergravity have a much better renormalization properties 
than the standard field theory. Simply, symmetry kills 
divergences but, only, an infinite dimensional symmetric model 
is the best choice. Superstring profited from this 
characteristic, over the string sheet, the theory posses 
superconformal symmetry \cite{gws}. In two dimensions (and only two), 
the conformal symmetry 
group is infinite dimensions. 
Working over $R^2$, we can 
define a single natural complex coordinate
\begin{eqnarray}
x^\mu = \{ x^1,x^2 \} \Longrightarrow z=x^1 + i x^2.
\end{eqnarray}
Any general conformal coordinate transformation in two 
dimensions is analytic in this complex coordinate.
\begin{eqnarray}
 z^{\prime} = f(z), \quad \overline{z}^{\prime} = 
\overline{f}(\overline{z}).
\end{eqnarray}
Owing to the Cauchy-Riemann condition;
\begin{eqnarray}
\frac{\partial f}{\partial x_0} = - e_1
\frac{\partial f}{\partial x_1},
\label{com}
\end{eqnarray}
which is, but not in a formal way \cite{ah}, equivalent to
\begin{eqnarray}
\frac{\partial}{\partial\overline{z}} f(z) = 0 .
\label{com1}
\end{eqnarray}

The quest of extending this notion of analyticity to four 
dimensions was studied extensively in the recent years. 
The most promising line of attack is quaternion analysis
\cite{gur1,ogv,dic}.
In this article, we want to investigate in details the 
standard generalization of complex analysis to the 
quaternionic case. We will be able to 
show that a new  generalization
of complex analysis to the quaternionic 
case indeed exists. The main novel result will be :
An 
infinite set of quaternionic analytic functions
obey our new analyticity condition.

\section{The Problem}
In four dimensions, quaternions are the natural extension of 
complex numbers. Over $R^4$, coordinates are well known to be 
unified into a quaternion
\begin{eqnarray}
x_\mu = \{ x_0,x_1,x_2,x_3 \} \Longrightarrow
 q =  x_0 + e_i x_i,
\label{q1}
\end{eqnarray}
where $e_i$ are the imaginary quaternion units 
 satisfying $e_i e_j = -\delta_{ij} + 
\epsilon_{ijk}e_k$ ($i,j,k=1,2,3$). The standard conjugation, involution
, is defined by
\begin{eqnarray}
\overline{q} = x_0 - e_i x_i.
\end{eqnarray}
But, unfortunately, the quaternion analysis has proved to be a 
very delicate subject.

In complete analogy to complex analysis, (following 
Ahlfors \cite{ah}), a quaternionic function 
\begin{eqnarray}
f(q) = f_0(x_{\mu}) e_0 + f_1(x_\mu) e_1
+ f_2(x_\mu) e_2 + f_3(x_\mu) e_3,
\end{eqnarray} 
is analytic if it 
obeys the following Generalized-Cauchy-Riemann condition
\begin{eqnarray}
\frac{\partial f(q)}{\partial x_{0}} = 
-e_{1} \frac{\partial f(q)}{\partial x_{1}} = 
-e_{2} \frac{\partial f(q)}{\partial x_{2}} =
-e_{3} \frac{\partial f(q)}{\partial x_3}.
\label{left}
\end{eqnarray}
The different four equalities ensures the same derivative
regardless of the direction of the approaching.
We have just met the first obstacle, namely the position of 
the $e_i$. Why not
\begin{eqnarray}
\frac{\partial f(q)}{\partial x_{0}} = 
- \frac{\partial f(q)}{\partial x_{1}} e_{1}= 
- \frac{\partial f(q)}{\partial x_{2}} e_{2}=
- \frac{\partial f(q)}{\partial x_{3}} e_{3}.
\label{right}
\end{eqnarray}
There is an ambiguity due to the non-commutativity, we will call  
(\ref{left}, \ref{right}) left and right quaternion analytic 
condition respectively.
But if we want to translate the four--dimensional physics to 
quaternionic language then which derivative should we use?
Actually, both of (\ref{left}) and  
(\ref{right}) are too restricted.
To show this, 
just  
plug $f(q)$ into (\ref{left}, \ref{right}), we get for the different imaginary 
units the following equalities, for the left case
\begin{eqnarray}
\frac{\partial f_0}{\partial x_0} =
\frac{\partial f_1}{\partial x_1} =
\frac{\partial f_2}{\partial x_2} =
\frac{\partial f_3}{\partial x_3}; \nonumber \\
\frac{\partial f_1}{\partial x_0} =
-\frac{\partial f_0}{\partial x_1} =
-\frac{\partial f_3}{\partial x_2} =
\frac{\partial f_2}{\partial x_3}; \nonumber \\
\frac{\partial f_2}{\partial x_0} =
\frac{\partial f_3}{\partial x_1} =
-\frac{\partial f_0}{\partial x_2} =
-\frac{\partial f_1}{\partial x_3};  \nonumber \\
\frac{\partial f_3}{\partial x_0} = 
-\frac{\partial f_2}{\partial x_1} =
\frac{\partial f_1}{\partial x_2} =
-\frac{\partial f_0}{\partial x_3},
\label{sys}
\end{eqnarray}
and for the right case
\begin{eqnarray}
\frac{\partial f_0}{\partial x_0} =
\frac{\partial f_1}{\partial x_1} =
\frac{\partial f_2}{\partial x_2} =
\frac{\partial f_3}{\partial x_3}; \nonumber \\
\frac{\partial f_1}{\partial x_0} =
-\frac{\partial f_0}{\partial x_1} =
\frac{\partial f_3}{\partial x_2} =
-\frac{\partial f_2}{\partial x_3}; \nonumber \\
\frac{\partial f_2}{\partial x_0} =
- \frac{\partial f_3}{\partial x_1} =
-\frac{\partial f_0}{\partial x_2} =
\frac{\partial f_1}{\partial x_3};  \nonumber \\
\frac{\partial f_3}{\partial x_0} = 
\frac{\partial f_2}{\partial x_1} =
-\frac{\partial f_1}{\partial x_2} =
-\frac{\partial f_0}{\partial x_3}.
\label{sys1}
\end{eqnarray}

We can put 
the common part of (\ref{sys}, \ref{sys1}) into the following form
\begin{eqnarray}
\frac{\partial^2 f_0(q)}{\partial x_0^{\ 2}} +
\frac{\partial^2 f_0(q)}{\partial x_1^{\ 2}} = 0, 
\quad
\frac{\partial^2 f_0(q)}{\partial x_0^{\ 2}} +
\frac{\partial^2 f_0(q)}{\partial x_2^{\ 2}} = 0,
\nonumber \\
\frac{\partial^2 f_0(q)}{\partial x_0^{\ 2}} +
\frac{\partial^2 f_0(q)}{\partial x_3^{\ 2}} = 0, 
\quad
\frac{\partial^2 f_0(q)}{\partial x_1^{\ 2}} +
\frac{\partial^2 f_0(q)}{\partial x_2^{\ 2}} = 0,
\nonumber \\
\frac{\partial^2 f_0(q)}{\partial x_1^{\ 2}} +
\frac{\partial^2 f_0(q)}{\partial x_3^{\ 2}} = 0,
\quad
\frac{\partial^2 f_0(q)}{\partial x_2^{\ 2}} +
\frac{\partial^2 f_0(q)}{\partial x_3^{\ 2}} = 0.
\end{eqnarray}
which leads eventually to 
\begin{eqnarray}
\frac{\partial^2 f_0(q)}{\partial x_0^{\ 2}} 
= 
- \frac{\partial^2 f_0(q)}{\partial x_1^{\ 2}} 
=
- \frac{\partial^2 f_0(q)}{\partial x_2^{\ 2}} 
=
- \frac{\partial^2 f_0(q)}{\partial x_3^{\ 2}} 
= 0,
\end{eqnarray}
and similar forms can be obtained for $f_1, f_2$ and $f_3$.
After considering the non-common terms, it is easy to see that
(\ref{sys},\ref{sys1}) admit only a special form of the linear function
as a solution. In complete disagreement with the complex case 
where analytic functions are wide class.

The only successful attempt to relax this 
constraint is due to 
Feuter \cite{gur1}, in analogy with 
(\ref{com1}), He defined the analyticity condition to be   
\begin{eqnarray}
\frac{\partial}{\partial\overline{q}} f(q) = 0, 
\end{eqnarray}
where 
\begin{eqnarray}
\frac{\partial}{\partial\overline{q}} \equiv
\frac{\partial}{\partial x_0} +
e_1 \frac{\partial}{\partial x_1} +
e_2 \frac{\partial}{\partial x_2} +
e_3 \frac{\partial}{\partial x_3} .
\end{eqnarray}
But, again, we have a problem due to the 
non-commutativity of quaternions.
Feuter defined left quaternionic analytic condition by
\begin{eqnarray}
 \stackrel{\longrightarrow}{
 \frac{\partial}{\partial\overline{q}}} f(q) 
&=&\frac{\partial}{\partial x_0} 
( f_0 + f_1 e_1 + f_2 e_2 + f_3 e_3 ) \nonumber \\
&+& e_1 \frac{\partial}{\partial x_1}
( f_0 + f_1 e_1 + f_2 e_2 + f_3 e_3 ) \nonumber \\
&+& e_2 \frac{\partial}{\partial x_2} 
( f_0 + f_1 e_1 + f_2 e_2 + f_3 e_3 ) \nonumber \\
&+& e_3 \frac{\partial}{\partial x_3} 
( f_0 + f_1 e_1 + f_2 e_2 + f_3 e_3 ) = 0, \label{lff}
\end{eqnarray}
and right quaternionic analytic condition by
\begin{eqnarray}
\stackrel{\longleftarrow}
{\frac{\partial}{\partial\overline{q}}} f(q) 
&=& \frac{\partial}{\partial x_0} 
( f_0 + f_1 e_1 + f_2 e_2 + f_3 e_3 ) \nonumber \\
&+&  \frac{\partial}{\partial x_1}
( f_0 + f_1 e_1 + f_2 e_2 + f_3 e_3 ) e_1 \nonumber \\
&+&  \frac{\partial}{\partial x_2}
( f_0 + f_1 e_1 + f_2 e_2 + f_3 e_3 ) e_2 \nonumber \\
&+&  \frac{\partial}{\partial x_3} 
( f_0 + f_1 e_1 + f_2 e_2 + f_3 e_3 ) e_3 = 0 .
\end{eqnarray}
After substituting by $f(q)$, we find, for the left 
analyticity condition
\begin{eqnarray}
& & \frac{\partial f_0}{\partial x_0} -
\frac{\partial f_1}{\partial x_1} -
\frac{\partial f_2}{\partial x_2} -
\frac{\partial f_3}{\partial x_3} = 0; \nonumber \\
& & \frac{\partial f_1}{\partial x_0} +
\frac{\partial f_0}{\partial x_1} +
\frac{\partial f_3}{\partial x_2} -
\frac{\partial f_2}{\partial x_3} = 0; \nonumber \\
& & \frac{\partial f_2}{\partial x_0} -
\frac{\partial f_3}{\partial x_1} +
\frac{\partial f_0}{\partial x_2} +
\frac{\partial f_1}{\partial x_3} = 0; \nonumber \\
& & \frac{\partial f_3}{\partial x_0} +
\frac{\partial f_2}{\partial x_1} -
\frac{\partial f_1}{\partial x_2} +
\frac{\partial f_0}{\partial x_3} = 0 ,
\end{eqnarray}
whereas for the right case, we get
\begin{eqnarray}
& & \frac{\partial f_0}{\partial x_0} -
\frac{\partial f_1}{\partial x_1} -
\frac{\partial f_2}{\partial x_2} -
\frac{\partial f_3}{\partial x_3} = 0; \nonumber \\
& & \frac{\partial f_1}{\partial x_0} +
\frac{\partial f_0}{\partial x_1} -
\frac{\partial f_3}{\partial x_2} +
\frac{\partial f_2}{\partial x_3} = 0; \nonumber \\
& & \frac{\partial f_2}{\partial x_0} +
\frac{\partial f_3}{\partial x_1} +
\frac{\partial f_0}{\partial x_2} -
\frac{\partial f_1}{\partial x_3} = 0; \nonumber \\
& & \frac{\partial f_3}{\partial x_0} -
\frac{\partial f_2}{\partial x_1} +
\frac{\partial f_1}{\partial x_2} +
\frac{\partial f_0}{\partial x_3} = 0 .
\end{eqnarray}
It is clear that these conditions are completely different
from the standard left/right analytic conditions defined in
(\ref{sys}, \ref{sys1}).
Such Feuter quaternionic analytic 
function admits right/left Weierstrass-like series
for left/right Feuter conditions respectively 
\cite{gur1}. But, it 
is not a straightforward generalization of complex analysis
as it is not based on (\ref{com}).
One can try simply
\begin{eqnarray}
f(q) &=& q^2 
= x_0^2 - x_1^2 - x_2^2 - x_3^2
+ 2 (e_1 x_1 + e_2 x_2  + e_3 x_3) x_0.
 \nonumber \\
\end{eqnarray}
It does not satisfy neither the left nor
the right Feuter analytic condition.

\section{the Solution}
The aim of this work is to find
a ring of analytic differentiable quaternionic function 
that generalizes the notion 
of complex analysis. The only way to relax the condition
(\ref{sys}, \ref{sys1}) is
to use both of left and right
derivative  together to define a new analytic condition.
The simplest case is
\begin{eqnarray}  
f^{\prime}(q) &=&
\frac{\partial f(q)}{\partial x_{0}} 
\nonumber \\
&=& - \frac{1}{2}
\left( e_{1} \frac{\partial f(q)}{\partial x_{1}} 
+ \frac{\partial f(q)}{\partial x_{1}} e_{1} \right) 
\nonumber \\ &=&
- \frac{1}{2}
\left( e_{2} \frac{\partial f(q)}{\partial x_{2}} 
+ \frac{\partial f(q)}{\partial x_{2}} e_{2} \right) 
\nonumber \\ &=&
- \frac{1}{2}
\left( e_{3} \frac{\partial f(q)}{\partial x_3}
+ \frac{\partial f(q)}{\partial x_{3}} e_{3} \right).
\label{genr1}
\end{eqnarray}
But unfortunately this 
simple generalization is also limited.
After substituting by $f(q)$, we get
\begin{eqnarray}
f^{\prime}(q) &=&
\frac{\partial f_0}{\partial x_0} 
+ \frac{\partial f_1}{\partial x_0} e_1
+ \frac{\partial f_2}{\partial x_0} e_2
+ \frac{\partial f_3}{\partial x_0} e_3
\nonumber \\
&=& \frac{\partial f_1}{\partial x_1} 
-  \frac{\partial f_0}{\partial x_1} e_1
\nonumber \\ 
&=& \frac{\partial f_2}{\partial x_2} 
-  \frac{\partial f_0}{\partial x_2} e_2
\nonumber \\ 
&=& \frac{\partial f_3}{\partial x_3} 
-  \frac{\partial f_0}{\partial x_3} e_3,
\end{eqnarray}
leading to 
\begin{eqnarray}
\frac{\partial f_0}{\partial x_0} =
\frac{\partial f_1}{\partial x_1} &=&
\frac{\partial f_2}{\partial x_2} =
\frac{\partial f_3}{\partial x_3}; \nonumber \\
\frac{\partial f_1}{\partial x_0} &=&
-\frac{\partial f_0}{\partial x_1} = 0 ; \nonumber \\
\frac{\partial f_2}{\partial x_0} &=&
-\frac{\partial f_0}{\partial x_2} = 0;  \nonumber  \\
\frac{\partial f_3}{\partial x_0} &=& 
-\frac{\partial f_0}{\partial x_3} = 0.
\end{eqnarray}
Which can work for some special forms but if one tries a simple function
like $q^2$, it will fail in contrast to the complex analysis.
So we should modify again this condition. 
After some trials, one can find the possible definition  
mimicking complex analysis is the following
\begin{eqnarray}
f^{\prime}(q) &=&
\frac{\partial f(q)}{\partial x_{0}} 
\nonumber \\
&=& - \frac{1}{2}
\left( e_{1} \frac{\partial f(q)}{\partial x_{1}} 
+ \frac{\partial f(q)}{\partial x_{1}} e_{1} \right) 
- \frac{\partial f_0}{\partial x_2} e_2
- \frac{\partial f_0}{\partial x_3} e_3
\nonumber \\ &=&
- \frac{1}{2}
\left( e_{2} \frac{\partial f(q)}{\partial x_{2}} 
+ \frac{\partial f(q)}{\partial x_{2}} e_{2} \right) 
- \frac{\partial f_0}{\partial x_1} e_1
- \frac{\partial f_0}{\partial x_3} e_3
\nonumber \\ &=&
- \frac{1}{2}
\left( e_{3} \frac{\partial f(q)}{\partial x_3}
+ \frac{\partial f(q)}{\partial x_{3}} e_{3} \right)
- \frac{\partial f_0}{\partial x_1} e_1
- \frac{\partial f_0}{\partial x_2} e_2,
\label{genr}
\end{eqnarray}
or  explicitly
\begin{eqnarray}  
f^{\prime}(q) &=&
\frac{\partial f_0}{\partial x_0} 
+ \frac{\partial f_1}{\partial x_0} e_1
+ \frac{\partial f_2}{\partial x_0} e_2
+ \frac{\partial f_3}{\partial x_0} e_3
\nonumber \\
&=& \frac{\partial f_1}{\partial x_1} 
- \frac{\partial f_0}{\partial x_1} e_1
- \frac{\partial f_0}{\partial x_2} e_2
- \frac{\partial f_0}{\partial x_3} e_3
\nonumber \\ 
&=& \frac{\partial f_2}{\partial x_2} 
- \frac{\partial f_0}{\partial x_1} e_1
- \frac{\partial f_0}{\partial x_2} e_2
- \frac{\partial f_0}{\partial x_3} e_3
\nonumber \\ 
&=& \frac{\partial f_3}{\partial x_3} 
- \frac{\partial f_0}{\partial x_1} e_1
- \frac{\partial f_0}{\partial x_2} e_2
- \frac{\partial f_0}{\partial x_3} e_3,
\end{eqnarray}
our analytic conditions are
\begin{eqnarray}
\frac{\partial f_0}{\partial x_0} =
\frac{\partial f_1}{\partial x_1} &=&
\frac{\partial f_2}{\partial x_2} =
\frac{\partial f_3}{\partial x_3}; \nonumber \\
\frac{\partial f_1}{\partial x_0} &=&
-\frac{\partial f_0}{\partial x_1} ; \nonumber \\
\frac{\partial f_2}{\partial x_0} &=&
-\frac{\partial f_0}{\partial x_2} ;  \nonumber  \\
\frac{\partial f_3}{\partial x_0} &=& 
-\frac{\partial f_0}{\partial x_3} .
\label{gqac}
\end{eqnarray}

We will call these conditions the Quaternionic 
Analyticity Condition (QAC). Moreover, we will call 
any quaternion function 
that satisfies the QAC simply Quaternionic Analytic 
Function (QAF). 
It is easy to show that 
(\ref{gqac}) can be put it in the following form
\begin{eqnarray}
\frac{\partial^2 f_0(q)}{\partial x_0^{\ 2}} +
\frac{\partial^2 f_0(q)}{\partial x_1^{\ 2}} = 0, 
\quad
\frac{\partial^2 f_0(q)}{\partial x_0^{\ 2}} +
\frac{\partial^2 f_0(q)}{\partial x_2^{\ 2}} = 0,
\nonumber \\
\frac{\partial^2 f_0(q)}{\partial x_0^{\ 2}} +
\frac{\partial^2 f_0(q)}{\partial x_3^{\ 2}} = 0, 
\quad
\frac{\partial^2 f_1(q)}{\partial x_0^{\ 2}} +
\frac{\partial^2 f_1(q)}{\partial x_1^{\ 2}} = 0,
\nonumber \\
\frac{\partial^2 f_2(q)}{\partial x_0^{\ 2}} +
\frac{\partial^2 f_2(q)}{\partial x_2^{\ 2}} = 0,
\quad
\frac{\partial^2 f_3(q)}{\partial x_0^{\ 2}} +
\frac{\partial^2 f_3(q)}{\partial x_3^{\ 2}} = 0.
\end{eqnarray}

Consider, again,
$
f(q) = q^2 
$.
It satisfies the QAC without any 
problems leading to 
\begin{eqnarray}
\frac{\partial f_0}{\partial x_0} =
\frac{\partial f_1}{\partial x_1} &=&
\frac{\partial f_2}{\partial x_2} =
\frac{\partial f_3}{\partial x_3} = 2 x_0 ; \nonumber \\
\frac{\partial f_1}{\partial x_0} &=&
-\frac{\partial f_0}{\partial x_1} = 2 x_1 ; \nonumber \\
\frac{\partial f_2}{\partial x_0} &=&
-\frac{\partial f_0}{\partial x_2} = 2 x_2 ;  \nonumber  \\
\frac{\partial f_3}{\partial x_0} &=& 
-\frac{\partial f_0}{\partial x_3} = 2 x_3,
\label{gqac1}
\end{eqnarray}
implying
\begin{eqnarray}
f^{\prime}(q) = 2q.
\end{eqnarray} 
And
\begin{eqnarray}
f_0^{\prime}(q) = 2 x_0 ; \nonumber \\
f_1^{\prime}(q) = 2 x_1 ; \nonumber \\
f_2^{\prime}(q) = 2 x_2 ; \nonumber \\
f_3^{\prime}(q) = 2 x_3 .
\end{eqnarray}
Also, we notice that
\begin{eqnarray}
\frac{\partial f_1}{\partial x_2} &=&
\frac{\partial f_1}{\partial x_3} = 0; \nonumber \\
\frac{\partial f_2}{\partial x_1} &=&
\frac{\partial f_2}{\partial x_3} = 0; \nonumber \\
\frac{\partial f_3}{\partial x_1} &=&
\frac{\partial f_3}{\partial x_2} = 0. 
\end{eqnarray}

One can prove without any problems that the sum  
of any two 
QAF is
again analytic.
Also 
\begin{eqnarray}
f(q) &=& f_0(x_{\mu}) e_0 + f_1(x_\mu) e_1
+ f_2(x_\mu) e_2 + f_3(x_\mu) e_3, \\
g(q) &=& g_0(x_{\mu}) e_0 + g_1(x_\mu) e_1
+ g_2(x_\mu) e_2 + g_3(x_\mu) e_3, \\
(fg)(q) &=&  (f_0 g_0 - f_1 g_1  - f_2 g_2 
-f_3 g_3 )(x_{\mu}) e_0 + 
(f_0 g_1 + f_1 g_0 + f_2 g_3 - f_3 g_2)(x_\mu) e_1
\nonumber \\
&+& (f_0 g_2 + f_2 g_0 + f_3 g_1 - f_1 g_3 )(x_\mu) e_2 + 
(f_0 g_3 + f_3 g_0 + f_1 g_2 - f_2 g_1) (x_\mu) e_3,\\
\end{eqnarray}  
taking into account $f_{0\ldots 3}, g_{0\ldots 3} 
\in {\cal R}$ then
after simple ,but lengthy, algebraic calculations
(Actually, the next relation holds trivially since 
$f^{\prime}(q) = \partial_0 f$ by \ref{genr}.) 
we can prove that 
\begin{eqnarray}
(fg)^{\prime} = f^{\prime} g + f g^{\prime},
\end{eqnarray}
But, in contrast to complex analysis, this does not
imply that every polynomial is QAF \footnote{
That $q^3$ is not analytic had been noticed by  
S. Abdel-Rahman and G. Auberson.}e.g, 
\begin{eqnarray}
q^3 = 
(q_0^3 - 3q_0 q_1^2 - 3 q_0 q_2^2 - 3 q_0 q_3^2 ) +
   (3 q_0^2 q_1 - q_1^3 - q_1 q_2^2 - q_1 q_3^2 ) e_1 \nonumber \\
+  (3 q_0^2 q_2 - q_2^3 - q_2 q_1^2 - q_2 q_3^2 ) e_2
+  (3 q_0^2 q_3 - q_3^3 - q_3 q_2^2 - q_3 q_1^2 ) e_3 ,
\end{eqnarray}
it is clear that the first condition of (\ref{gqac}) is no
more valid. the possible solution is the following modification
\footnote{This suggestion is entirely due to S. Abdel-Rahman.}
(no summation over i=1,2,3)
\begin{eqnarray}
\frac{\partial f_0}{\partial x_0} &=&
{\frac{1}{x_i}}
\left( x_1 \frac{\partial }{\partial x_1} 
+ x_2 \frac{\partial }{\partial x_2} 
+ x_3 \frac{\partial }{\partial x_3}
\right) f_i; \nonumber \\
\frac{\partial f_1}{\partial x_0} &=&
-\frac{\partial f_0}{\partial x_1} ; \nonumber \\
\frac{\partial f_2}{\partial x_0} &=&
-\frac{\partial f_0}{\partial x_2} ;  \nonumber  \\
\frac{\partial f_3}{\partial x_0} &=& 
-\frac{\partial f_0}{\partial x_3} .
\label{mgqac}
\end{eqnarray}
I will call this condition the Modified QAF (MQAF), which 
is valid for any polynomial, (for integer n and $a_{0 \ldots n} 
\in {\cal R}$) 
\begin{eqnarray}
P(q) = a_0  + a_1 q + \ldots + a_n q^n,
\end{eqnarray}
is a MQAF and
\begin{eqnarray}
P^{\prime}(q) = a_1 + \ldots + n a_n q^{n-1}.
\end{eqnarray}
In complete parallel agreement with complex analysis.
In summary (\ref{gqac}) admits a chain rule,
 it has a simple geometric meaning something like the 
overlapping of three planar complex structure, in the spirit 
of \cite{ansfre}, a tri-holomorphicity that appears in the context
of Self-Dual Yang-Mills fields. But being non-valid 
for generic polynomials, we can not develop something as 
powerful as Laurent series whereas (\ref{mgqac}) is more attractive 
and may have a richer geometric structure.
I have given in the appendix a Mathematica\cite{math} program to check
(\ref{mgqac}) to any desired order.

\section{Further Extension}
The next step should be octonions. Every thing (about the QAC) is the same 
just we let the i index takes values from 1 to 7.
Our Octonionic Analytic Condition OAC reads
(where summation over repeated indices is understood)
\begin{eqnarray}  
f^{\prime}(q) &=&
\frac{\partial f_0}{\partial x_0} 
+ \frac{\partial f_i}{\partial x_0} e_i
= \frac{\partial f_1}{\partial x_1} 
- \frac{\partial f_0}{\partial x_i} e_i
\nonumber \\ 
&=& \frac{\partial f_2}{\partial x_2} 
- \frac{\partial f_0}{\partial x_i} e_i
= \frac{\partial f_3}{\partial x_3} 
- \frac{\partial f_0}{\partial x_i} e_i
\nonumber \\
&=& \frac{\partial f_4}{\partial x_4} 
- \frac{\partial f_0}{\partial x_i} e_i
= \frac{\partial f_5}{\partial x_5} 
- \frac{\partial f_0}{\partial x_i} e_i
\nonumber \\ 
&=& \frac{\partial f_6}{\partial x_6} 
- \frac{\partial f_0}{\partial x_i} e_i
= \frac{\partial f_7}{\partial x_7} 
- \frac{\partial f_0}{\partial x_i} e_i
\end{eqnarray}
which implies
\begin{eqnarray}
\frac{\partial f_0}{\partial x_0} =
\frac{\partial f_1}{\partial x_1} =
\frac{\partial f_2}{\partial x_2} =
\frac{\partial f_3}{\partial x_3} &=&
\frac{\partial f_4}{\partial x_4} =
\frac{\partial f_5}{\partial x_5} =
\frac{\partial f_6}{\partial x_6} =
\frac{\partial f_7}{\partial x_7}; \nonumber \\
\frac{\partial f_1}{\partial x_0} &=&
-\frac{\partial f_0}{\partial x_1} ; \nonumber \\
\frac{\partial f_2}{\partial x_0} &=&
-\frac{\partial f_0}{\partial x_2} ; \nonumber \\
\frac{\partial f_3}{\partial x_0} &=& 
-\frac{\partial f_0}{\partial x_3} ; \nonumber \\
\frac{\partial f_4}{\partial x_0} &=&
-\frac{\partial f_4}{\partial x_4} ; \nonumber  \\
\frac{\partial f_5}{\partial x_0} &=&
-\frac{\partial f_5}{\partial x_5} ; \nonumber  \\
\frac{\partial f_6}{\partial x_0} &=&
-\frac{\partial f_6}{\partial x_6} ; \nonumber  \\
\frac{\partial f_7}{\partial x_0} &=&
-\frac{\partial f_7}{\partial x_7} ,
\label{goac}
\end{eqnarray}
or in the simple form
\begin{eqnarray}
\frac{\partial^2 f_0(q)}{\partial x_0^{\ 2}} +
\frac{\partial^2 f_0(q)}{\partial x_1^{\ 2}} = 0, 
\quad
\frac{\partial^2 f_0(q)}{\partial x_0^{\ 2}} +
\frac{\partial^2 f_0(q)}{\partial x_2^{\ 2}} = 0,
\nonumber \\
\frac{\partial^2 f_0(q)}{\partial x_0^{\ 2}} +
\frac{\partial^2 f_0(q)}{\partial x_3^{\ 2}} = 0, 
\quad
\frac{\partial^2 f_0(q)}{\partial x_0^{\ 2}} +
\frac{\partial^2 f_0(q)}{\partial x_4^{\ 2}} = 0,
\nonumber \\
\frac{\partial^2 f_0(q)}{\partial x_0^{\ 2}} +
\frac{\partial^2 f_0(q)}{\partial x_5^{\ 2}} = 0, 
\quad
\frac{\partial^2 f_0(q)}{\partial x_0^{\ 2}} +
\frac{\partial^2 f_0(q)}{\partial x_6^{\ 2}} = 0,
\nonumber \\
\frac{\partial^2 f_0(q)}{\partial x_0^{\ 2}} +
\frac{\partial^2 f_0(q)}{\partial x_7^{\ 2}} = 0, 
\quad
\frac{\partial^2 f_1(q)}{\partial x_0^{\ 2}} +
\frac{\partial^2 f_1(q)}{\partial x_1^{\ 2}} = 0,
\nonumber \\
\frac{\partial^2 f_2(q)}{\partial x_0^{\ 2}} +
\frac{\partial^2 f_2(q)}{\partial x_2^{\ 2}} = 0,
\quad
\frac{\partial^2 f_3(q)}{\partial x_0^{\ 2}} +
\frac{\partial^2 f_3(q)}{\partial x_3^{\ 2}} = 0.
\nonumber \\
\frac{\partial^2 f_4(q)}{\partial x_0^{\ 2}} +
\frac{\partial^2 f_4(q)}{\partial x_4^{\ 2}} = 0.
\quad
\frac{\partial^2 f_5(q)}{\partial x_0^{\ 2}} +
\frac{\partial^2 f_5(q)}{\partial x_5^{\ 2}} = 0.
\nonumber \\
\frac{\partial^2 f_6(q)}{\partial x_0^{\ 2}} +
\frac{\partial^2 f_6(q)}{\partial x_6^{\ 2}} = 0.
\quad
\frac{\partial^2 f_7(q)}{\partial x_0^{\ 2}} +
\frac{\partial^2 f_7(q)}{\partial x_7^{\ 2}} = 0.
\end{eqnarray}
The results obtained for quaternionic QAC can be generalized 
directly to octonions without any problems since the 
non-associativity of octonions is not relevant here.
But generalizing the MQAF to octonions is not at all easy
since, starting from $f^3$, any octonionic polynomial is ill defined
\begin{equation}
f^3(x_\mu)\ \quad \mbox{is it} \ \quad (f.f).f\ \quad \mbox{or} \ 
\quad f.(f.f),
\end{equation} 
and the number of alternatives 
increases with the power of $f$. It would be interesting
to find out if a possible combination admits something
like the MQAC.
\section{Conclusion}
To conclude, we have 
found a natural extension of Cauchy--Reimann condition
to 4  which we hope to play 
the role of complex analysis in 2 dimensions.
It will be interesting to look for a generalization of
conformal symmetry and  
Virasoro algebra \cite{gws} which will lead us eventually 
to find integrable models in higher dimensions.
The case of octonions is still mysterious.
\section*{Acknowledgements}

I would like to acknowledge many fruitful discussions
with P. Rotelli  as well as the physics department
at Lecce university for their kind hospitality.
I am grateful to Prof. A.~Zichichi 
and the ICSC--World Laboratory for financial
support. Lastly, I am indebted to S.~Abdel-Rahman, 
D.~Anselmi and G.~Auberson for their very useful comments.

\appendix
\section{}

I have ran this program to $f^{10}$. Since the presence of
a chain rule is not enough to guarantee the analyticity of generic 
polynomials, this seems to be the only way - Explicit Calculation.
Below, I give the programm and an example the test of analyticity
of $f^{10}$\\
e0 = \{\{1,0\},\{0,1\}\};
\\e1 = \{\{0,-I\},\{-I,0\}\};
\\e2 = \{\{0,-1\},\{1,0\}\};
\\e3 = \{\{-I,0\},\{0,I\}\};
\\Q = q0 e0 + q1 e1 + q2 e2 + q3 e3;
\\f0[x\_] := Expand[ ComplexExpand[
                                 (  ComplexExpand[
                                                   Conjugate[ Part[x,1,1]]
                                                  ] +
                                     Part[x,1,1]
                                  )/2
                                 ]
                    ];
\\f2[x\_] := Expand[ ComplexExpand[ ( ComplexExpand[ 
   Conjugate[ Part[x,2,1]] ] + 
          Part[x,2,1] )/2 ] ];
\\f1[x\_] := Expand[ ComplexExpand[ ( ComplexExpand[
   Conjugate[ Part[x,2,1]] ] - 
          Part[x,2,1] )/(  2 I) ] ];
\\f3[x\_] := Expand[ ComplexExpand[ ( ComplexExpand[
   Conjugate[ Part[x,1,1]] ] - 
          Part[x,1,1] )/(2 I) ] ];
\\d0[x\_,y\_] := Expand[
                      ComplexExpand[ 
                         ( q1 D[x,q1] + q2 D[x,q2] + q3 D[x,q3]) /y
                      ]
              ];
\\Q10=  Expand[Q.Q.Q.Q.Q.Q.Q.Q.Q.Q];
\\Print["10.........."];
\\Print[Expand[D[f0[Q10],q0]] == d0[f1[Q10],q1]];
\\Print[Expand[D[f0[Q10],q0]] == d0[f2[Q10],q2]];
\\Print[Expand[D[f0[Q10],q0]] == d0[f3[Q10],q3]];
\\Print[Expand[D[f0[Q10],q1]] == - Expand[D[f1[Q10],q0]]];
\\Print[Expand[D[f0[Q10],q2]] == - Expand[D[f2[Q10],q0]]];
\\Print[Expand[D[f0[Q10],q3]] == - Expand[D[f3[Q10],q0]]];
\\Print["Ready"];

\vspace{2cm}

The output is 
\\10..........
\\True
\\True
\\True
\\True
\\True
\\True
\\Ready
\\

\begin{thebibliography}{00}
\bibitem{zee} G. G. Ross, Grand Unified Theories, 
Reading: Benjamin/Cummings, 1985.
\bibitem{gws} M. B. Green, J. H. Schwarz and E. Witten, 
Superstring Theory, Cambridge University Press, 
1987.
\bibitem{ah} L. Ahlfors, Complex analysis, McGraw-Hill, 
International Edition 1979, page 25-27.
\bibitem{gur1} F. G\"{u}rsey and H. C. Tze,
``Complex and Quaternionic Analyticity in Chiral Gauge 
Theories, Part I'',  Annals of Physics, 128 (1978) 29.
\bibitem{ogv} M. Evans, F. G\"{u}rsey and V. Ogievetsky, 
``From 2D conformal to 4D self-dual theories: quaternionic 
analyticity'', CERN-TH.6533/92, hep-th/9207089.
\bibitem{dic} R. Dick, ``Half--Differentials and fermions'',
IASSNS-HEP-94/83, hep-th/9410099.
\bibitem{ansfre} D.~Anselmi~and~P.~Fr\`e, Nucl.~Phys.  B416  
(1994) 255.
\bibitem{math} S.~Wolfram, Mathematica, Addison-Wesley Publishing Company, 
Inc., Second Edition 1991.
\end{thebibliography}
\end{document}